\documentclass[12pt,preprint]{aastex}

\begin{document}

\shorttitle{A catalogue of $z>5.9$ Lyman break  galaxies in the WFC\,3-HUDF}
\title{A Matched Catalogue of  $z> 5.9$ Galaxies in the
WFC\,3 Hubble Ultra Deep Field}
\shortauthors{Bunker \& Wilkins}
\author{Andrew J.\ Bunker and Stephen M.\ Wilkins}
\affil{Department of Physics University of Oxford, Denys Wilkinson Building,
Keble Road, OX1\,3RH, U.K.\\ {\tt email:
a.bunker1@physics.ox.ac.uk}}

\begin{abstract}
There have been several independent analyses of the recent Wide Field Camera 3
images of the Hubble Deep Field, selecting galaxies at $z>6$ through
the Lyman break technique. Presented here is a matched catalogue of
objects in common between the analyses posted to this preprint server,
listing the different catalogue names associated with the same sources. 
\end{abstract}

In this brief paper, we collate several different catalogues of Lyman break
galaxies in the Wide Field Camera 3 imaging of the Hubble Ultra Deep Field
which have appeared since the data become public in September 2009.
We also cross-match sources with previously-published catalogues based
on the original Hubble Ultra Deep Field images with the Advanced Camera for
Surveys in 2004.

Table~1 is the list of 22 galaxies with photometric redshifts between $z=5.9$
and $z=6.4$ presented in McLure et al.\ (2009), with corresponding ID numbers
from Bunker et al.\ (2004) and Yan \& Windhorst (2004). Of the 22 galaxies
in this redshift range presented in the recent McLure et al.\ paper, 19 are
 $i'$-drops first identified in the Advanced Camera for Surveys
imaging of the Hubble Ultra Deep Field in 2004. Two of the remaining three are
very faint in the $z'$-band (with $z'_{AB}=29.24$ for object 2719 in McLure
et al., and $z'_{AB}=29.37$ for object 2003), and so were not selected in
$z$-band limited samples.
\begin{table*}
\begin{tabular}{rccccc}
\hline
{Original ID} & {McLure ID} & {RA(J2000)} & {Dec(J2000)} & {$z_{phot}$ (McLure)}\\
\hline
B\,46223 &  1735  & 03 32 39.86 & $-$27 46 19.1 &    5.90 &   \\
B\,16258 &  1955  & 03 32 39.46 & $-$27 45 43.4 &    5.90    \\
B\,41918 &   1719  & 03 32 44.70 & $-$27 46 45.6 &    5.95 &     \\
YW\,57&  2217  & 03 32 40.56 & $-$27 48 02.7 &    5.95 &      \\
B\,27270 &  962   & 03 32 35.05 & $-$27 47 40.1 &    5.95 &    \\
B\,49701 &  1189  & 03 32 36.98 & $-$27 45 57.6 &    6.00 &    \\
YW\,51 &  2830  & 03 32 34.58 & $-$27 46 58.0 &    6.00 &      \\
YW\,91 &  2498  & 03 32 35.04 & $-$27 47 25.8 &    6.00 &    \\
--- & 2719  & 03 32 40.59 & $-$27 45 56.9 &    6.05 & \\
B\,45467 &  1625  & 03 32 43.03 & $-$27 46 23.8 &    6.05 &   \\
B\,24733 &  1398  & 03 32 36.63 & $-$27 47 50.1 &    6.10 &   \\
YW\,52 &  1760  & 03 32 40.25 & $-$27 46 05.2 &    6.15 &   \\
B\,44194 &   934   & 03 32 37.48 & $-$27 46 32.5 &    6.20 &    \\
YW\,27 &  2791  & 03 32 36.64 & $-$27 47 50.2 &    6.25 &    \\
YW\,85 &  1464  & 03 32 42.19 & $-$27 46 27.9 &    6.30 &    \\
--- &  2003  & 03 32 36.46 & $-$27 47 32.4 &    6.30   &  \\
YW\,82 &  2514  & 03 32 39.79 & $-$27 46 33.8 &    6.30 &    \\
B\,44154 &  837   & 03 32 37.46 & $-$27 46 32.8 &    6.35 &    \\
YW\,95 &  1855  & 03 32 43.79 & $-$27 46 33.8 &    6.40 &  \\
YW\,56a &  1864  & 03 32 34.52 & $-$27 47 34.8 &    6.40 &     \\
YW\,75 (UDFz-36777536) &  1911            & 03 32 36.77 & $-$27 47 53.6 &    6.40 & \\
(UDFz-39586565) & 1915            & 03 32 39.58 & $-$27 46 56.5 &    6.40  &  \\
\hline\end{tabular}
\caption{The prefix `B' in the column 1 indicates a galaxy first identified as a candidate
$z\approx 6$ $i'$-band dropout by Bunker et al.\ (2004); the `YW' prefix indicates other
(typically fainter) $i'$-drops identified by Yan \& Windhorst (2004). McLure's objects 1911 \&
1915 have also recently been identified by Oesch et al.\ (2009), with this ID given in parentheses
in column 1.}
\end{table*}

 Table~2 presents the
McLure et al.\ sources with $z_{phot}>6.4$ with corresponding ID numbers from
Bunker et al.\ (2009), Oesch et al.\ (2009) and Bouwens et al.\ (2009).
There are 16 $z'$-drops in the Oesch et al.\ sample, compared with 10 from Bunker et al.
In all cases this is due to a slightly different colour selection of $(z'-Y)_{AB}>1$ in Bunker et al.\
compared with a bluer cut of $(z'-Y)_{AB}>0.8$ (which picks up more galaxies at lower
redshifts) or the brighter magnitude limit applied in Bunker et al.\ ($Y_{AB}<28.5$)
The Oesch et al.\ galaxies UDFz-36777536, UDFz-39586565 (in Table 1) have colours $0.8<(z'-Y)_{AB}<1$
and are likely to be at $z\approx 6.4$, and UDFz-37446513 also has colours too blue for
the Bunker et al.\ $z'$-drop selection. The Oesch et al.\ galaxies
UDFz-37807405 
and UDFz-41057156 
lie close to or fainter than $Y_{AB}=28.5$ and hence do not appear in the list of Bunker et al.
All the Oesch et al.\ galaxies appear in the list of McLure et al.\ except for 
UDFz-38537519 (03:32:38.53, $-$27:47:51.9, J2000) which is the faintest $z'$-drop in their
sample ($Y_{AB}=29.17$, $J_{AB}=29.13$) and is also too faint for the Bunker et al.\ selection.

For the $Y$-band drop-outs, all five galaxies presented in Bouwens et al.\ (2009) also appear
in Bunker et al.\ (2009) with the exception of UDFy-37796000, which is too faint for the Bunker
et al.\ cut of $J_{AB}<28.5$. Bunker et al.\ have three additional sources not appearing in
Bouwens et al., one of which (YD4 in Table 2) is also in the McLure et al.\ (2009) catalogue,
and two others (YD5 at 03:32:35.85 $-$27:47:17.1 \& YD6 at 03:32:40.40 $-$27:47:18.8)
absent from the other catalogues.

There is one source in McLure et al.\ (2009), object 1422, which lies within $0\farcs4$
of object zD4/UDFz-39557176/1092, and presumably is classed as a single object
in the other catalogues. There are 8 remaining sources in McLure et al.\ which do
not appear in the other catalogues, either because of their faint magnitudes or
their colours falling outside the selection windows for $z'$-drops and $Y$-drops.
We note that McLure et al.\ adopt an SED-fitting approach, and the other groups
utilize colour cuts.

  \begin{table*}
\begin{tabular}{rcccccc}
\hline
{Bunker ID} & {Oesch/Bouwens ID} & {McLure ID} & {RA(J2000)} & {Dec(J2000)} & {$z_{phot}$ (McLure)}\\
  \hline
--- &  --- & 2195  & 03 32 43.05 & $-$27 47 08.1 &    6.45  &  \\
--- & UDFz-37446513 &1880            & 03 32 37.44 & $-$27 46 51.3 &    6.50 &    \\
zD6 &  UDFz-36387163 & 1958            & 03 32 36.38 & $-$27 47 16.2 &    6.50 &  \\
zD8 &  UDFz-40566437 & 2206            & 03 32 40.58 & $-$27 46 43.6 &    6.50 &      \\
--- & --- & 1064  & 03 32 34.93 & $-$27 47 01.3 &    6.65 & \\
zD1 &  UDFz-42566566 & 688             & 03 32 42.56 & $-$27 46 56.6 &    6.70 &   \\
--- & --- & 2794  & 03 32 36.75 & $-$27 46 48.2 &    6.75  &  \\
zD3 &  UDFz-42577314 & 1144            & 03 32 42.56 & $-$27 47 31.5 &    6.80 &   \\
--- & --- & 2395  & 03 32 44.31 & $-$27 46 45.2 &    6.80  & \\
zD4 &  UDFz-39557176 & 1092            & 03 32 39.55 & $-$27 47 17.5 &    6.85 &    \\
--- & UDFz-37807405 & 2560            & 03 32 37.80 & $-$27 47 40.4 &    6.90 &    \\
--- & --- & 2826  & 03 32 37.06 & $-$27 48 15.2 &    6.90  &  \\
zD5 &  UDFz-43146285 & 1678            & 03 32 43.14 & $-$27 46 28.6 &    7.05 &     \\
zD10 & UDFz-39736214 & 2502            & 03 32 39.73 & $-$27 46 21.4 &    7.10 &    \\
zD9 &  UDFz-37228061 & 1574            & 03 32 37.21 & $-$27 48 06.1 &    7.20 &    \\
zD2 &  UDFz-38807073 & 835             & 03 32 38.81 & $-$27 47 07.2 &    7.20 &   \\
--- &  UDFz-41057156 & 2066            & 03 32 41.05 & $-$27 47 15.6 &    7.20 &    \\
--- & --- & 2888  & 03 32 44.75 & $-$27 46 45.1 &    7.35 &  \\
--- & --- & 2940  & 03 32 36.52 & $-$27 46 41.9 &    7.40   &\\
YD7 &  UDFy-37636015 & 2079            & 03 32 37.63 & $-$27 46 01.5 &    7.50 &    \\
zD7 &  UDFz-44716442 & 1107            & 03 32 44.71 & $-$27 46 44.4 &    7.60 &     \\
--- & ---  &1422$^{\dagger}$  & 03 32 39.52 & $-$27 47 17.3 &    7.60 &   \\
YD4 & --- & 2487  & 03 32 33.13 & $-$27 46 54.4 &    7.80 &    \\
YD1 & UDFy-42886345 & 1765            & 03 32 42.88 & $-$27 46 34.6 &    7.95 &   \\
--- &  UDFy-43086276 & 2841            & 03 32 43.09 & $-$27 46 27.9 &    8.10 &     \\
YD2 &  UDFy-37796000 & 1939            & 03 32 37.80 & $-$27 46 00.1 &    8.35 &    \\
YD3 & UDFy-38135539 & 1721            & 03 32 38.14 & $-$27 45 54.0 &    8.45 &     \\
\hline\end{tabular}

$^{\dagger}$ this object is within $0\farcs4$ of zD4/UDFz-39557176/1092.
\caption{The galaxies identified by McLure et al.\ (2009) which have photmetric redshifts $z>6.4$,
with the corresponding ID numbers from Bunker et al.\ (2009, column 1), and Oesch et al.\ (2009, prefix UDFz-)
or Bouwens et al.\ (2009, prefix UDFy-) in column 2.}
\end{table*}

\section*{References}

Bouwens, R.~J., Illingworth, G.~D., Oesch, P.~A.; Stiavelli, M., van Dokkum, P., Trenti, M., Magee, D., Labbe, I., Franx, M., Carollo, M.\
2009, arXiv:0909.1803 

Bunker, A.~J., Stanway, 
E.~R., Ellis, R.~S., \& McMahon, R.~G.\ 2004, MNRAS, 355, 374 

Bunker, A., Wilkins, S., Ellis, R., Stark, D., Lorenzoni, S., Chiu, K., Lacy, M., Jarvis, M., Hickey, S.\ 
2009, arXiv:0909.2255 

McLure, R.~J., Dunlop, 
J.~S., Cirasuolo, M., Koekemoer, A.~M., Sabbi, E., Stark, D.~P., Targett, 
T.~A., \& Ellis, R.~S.\ 2009, arXiv:0909.2437 

Oesch, P.~A., Bouwens, R.~J., Illingworth, G.~D., Carollo, C.~M., Franx, M., Labbe, I., Magee, D.,
 Stiavelli, M., Trenti, M., van Dokkum, P.~G.\ 2009, arXiv:0909.1806 
 
 Yan, H., \& Windhorst, R.~A.\ 2004, ApJ, 600, L1

\end{document}